\documentstyle[natbib,epsfig,longtable]{aipproc}

\newcommand{\nubar}[0]{\overline{\nu}}

\newcommand{\Vcd}{\rm |V_{cd}|}
\newcommand{\Vcs}{\rm |V_{cs}|}
\newcommand{\Vub}{\rm |V_{ub}|}
\newcommand{\Vcb}{\rm |V_{cb}|}
\newcommand{\Vtd}{\rm |V_{td}|}
\newcommand{\Vts}{\rm |V_{ts}|}
\newcommand{\Vtb}{\rm |V_{tb}|}
\newcommand{\Vud}{\rm |V_{ud}|}
\newcommand{\Vus}{\rm |V_{us}|}
\newcommand{\Vij}{\rm |V_{ij}|}
\newcommand{\pt}{\rm p_t}

\newcommand{\stw}{\mbox{$\sin^2\theta_W$}}
\newcommand{\ee}{{\rm e^+e^-}}

\newcommand{\bbar}{\overline{{\rm b}}}
\newcommand{\cbar}{\overline{{\rm c}}}
\newcommand{\bbbar}{{\rm b}\bbar}
\newcommand{\bcbar}{{\rm b}\cbar}
\newcommand{\cbbar}{{\rm c}\bbar}
\newcommand{\Enu}{{\rm E}_\nu}

\newcommand{\ECoM}{{\rm E_{CoM}}}

\begin{document}

\setcounter{secnumdepth}{2}    

\pagestyle{plain}
\footskip 1.5 cm

\title{Mighty MURINEs: Neutrino Physics at Very High Energy Muon Colliders
\thanks{
To appear in Proc. HEMC'99 Workshop -- Studies on Colliders and
Collider Physics at the Highest Energies: Muon Colliders at 10 TeV to
100 TeV; Montauk, NY, September 27-October 1, 1999,
http://pubweb.bnl.gov/people/bking/heshop/ .
This work was performed under the auspices of
the U.S. Department of Energy under contract no. DE-AC02-98CH10886.}
}

\author{Bruce J. King}
\address{Brookhaven National Laboratory\\
email: bking@bnl.gov\\
web page: http://pubweb.bnl.gov/people/bking}
\maketitle

\begin{abstract}
  An overview is given of the potential for neutrino physics studies
through parasitic use of the intense high energy neutrino
beams that would be produced at future many-TeV muon colliders.
Neutrino experiments clearly cannot compete with the collider physics.
Except at the very highest energy muon colliders, the main thrust of
the neutrino physics program would be
to improve on the measurements from preceding neutrino experiments
at lower energy muon colliders, particularly in the fields of B physics,
quark mixing and CP violation. Muon colliders at the 10 TeV energy scale
might already produce of order $10^8$ B hadrons per year in a favorable
and unique enough experimental environment to have some analytical capabilities
beyond any of the currently operating or proposed B factories. The most
important of the quark mixing measurements at these energies might well
be the improved measurements of the important CKM matrix elements $\Vub$
and $\Vcb$ and, possibly, the first measurements of $\Vtd$ in the
process of flavor
changing neutral current interactions involving a top quark loop. Muon
colliders at the highest center-of-mass energies that have been conjectured,
100--1000 TeV, would produce
neutrino beams for neutrino-nucleon interaction experiments
with maximum center-of-mass energies from 300--1000 GeV.
Such energies are
close to, or beyond, the discovery reach of all colliders before
the turn-on of the LHC. In particular, they are
comparable to the 314 GeV center-of-mass energy for
electron-proton scattering at the currently operating
HERA collider and so HERA provides a convenient benchmark for
the physics potential. It is shown that these ultimate
terrestrial neutrino experiments, should they eventually
come to pass, would have several orders of magnitude more
luminosity than HERA. This would potentially open up the possibility
for high statistics studies of any exotic particles, such as
leptoquarks, that might have been previously discovered
at these energy scales.
\end{abstract}

\section{Introduction: the Role of Mighty MURINES}
\label{sec:intro}

  The dominant motivation for high energy muon colliders (HEMCs)
is unquestionably to explore elementary particle physics at
many-TeV energy scales. For the sake of completeness, however,
this paper instead discusses what would be the most promising subsidiary
fixed target physics program, namely, the parasitic use of the
free and profuse neutrino beams at HEMCs to provide complementary
precision studies of high energy physics (HEP) at lower energies.
Perhaps, this might complement collider studies in fostering new
and helpful insights into the properties of elementary particles.

  Neutrino interactions
have unique potential for precision HEP studies because they only
participate in the weak interaction.
Today's
neutrino beams from pion decays lack the intensity to fully exploit
this potential but future MUon RIng Neutrino Experiments (MURINEs), using
neutrino beams from the decays of muons in a muon collider or other
storage ring, hold the
promise of neutrino beams that are several orders of magnitude more
intense than today's beams~\cite{bjkthesispaper}.
The first MURINEs may well be muon storage rings dedicated to
neutrino production~\cite{geer} (``neutrino factories'')
while the collider rings of any first generation muon colliders
will also make excellent MURINEs.
The topic of this paper is MURINEs at very high energies and
these will be referred
to as ``Mighty MURINEs''~\footnote{this is a play on words
alluding to the venerable cartoon character Mighty Mouse through
the dictionary definition of ``murine'' as ``to do with mice''.}.

  At a minimum, Mighty MURINEs will improve on previous MURINEs in
providing much useful bread-and-butter precision HEP to feed
the hungry masses of HEP experimentalists with aversions
to collider mega-experiments. More interestingly, there are a couple of
plausible scenarios under which they might do much more; namely (i) if
quark mixing and/or B physics offer more than predicted by our naive prejudices
as parameterized in the standard model (SM) of elementary particles, and
(ii) if leptoquarks or other exotica begin to emerge at or below the
100 GeV energy scale.

 The following section surveys the experimental conditions and
parameters that might be found at Mighty MURINEs. This is
followed by an overview of the potential physics analyses
and by three sections going into more detail on the most
interesting topics: one each on exploiting Mighty MURINEs as
B factories, on the possibilities for quark
mixing studies and on the potential for heavy particle production
up to the 100 GeV scale.

\section{Experimental Overview}
\label{sec:expt}

\subsection{The Neutrino Beams}
\label{subsec:expt_beam}

  Neutrinos are emitted from the decay of muons in the collider ring:
\begin{eqnarray}
\mu^- & \rightarrow & \nu_\mu + \overline{\nu_{\rm e}} + {\rm e}^-,
                                             \nonumber \\
\mu^+ & \rightarrow & \overline{\nu_\mu} + \nu_{\rm e} + {\rm e}^+.
                                                 \label{eq:nuprod}
\end{eqnarray}

\begin{figure}[t!] %
\centering
\includegraphics[height=2.5in,width=6.0in]{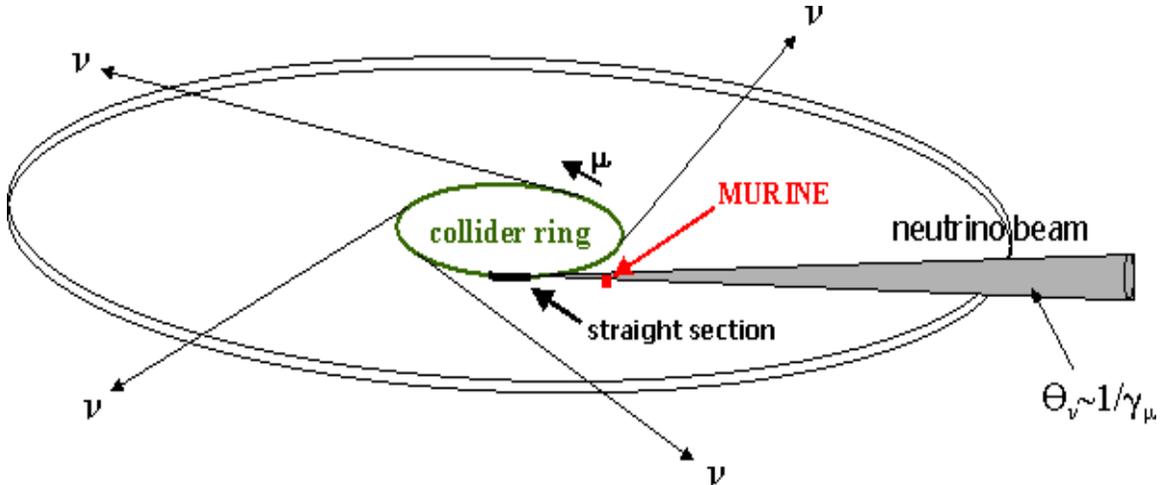}
\caption{
The decays of muons in a muon collider will produce a disk
of neutrinos emanating out tangentially from the collider ring.
The neutrinos from decays in straight sections will line up
into beams suitable for experiments. The MURINEs will be
sited in the center of the most intense beam and as close
as is feasible to the production straight section.
}
\label{nu_beam}
\end{figure}

  As is illustrated in figure~\ref{nu_beam},
the thin pencil beams of neutrinos for experiments will be produced
from the most suitable long straight sections in the collider ring or,
possibly, in the accelerating rings. These will be referred to as
the production straight sections.
The divergence of the neutrino beam is typically
dominated by the decay opening angles of the neutrinos rather than the
divergence of the parent muon beam. Relativistic kinematics boosts
the forward hemisphere
in the muon rest frame into a narrow cone in the laboratory
frame with a characteristic opening half-angle,
$\theta_\nu$,
given in obvious notation by
\begin{equation}
\theta_\nu \simeq \sin \theta_\nu = 1/\gamma_\mu =
\frac{m_\mu c^2}{E_\mu} \simeq \frac{10^{-4}}{E_\mu [{\rm TeV}]}.
                                                   \label{eq:thetanu}
\end{equation}
For the example of 5 TeV muons, the neutrino beam will have an opening
half-angle of approximately 0.02 mrad.

  The large muon currents and tight collimation of the neutrinos results
in such intense neutrino beams that potential radiation
hazards~\cite{bjkthesispaper,hemc99nurad}
are a serious design issue for the neutrino beam-line and even for
the less intense neutrino fluxes emanating from the rest of the collider
ring.

\subsection{Luminosities at Neutrino Experiments}
\label{subsec:expt_lum}

For a cylindrical experimental target extending out from the beam
center to an angle $\theta_\mu = 1/\gamma_\mu$,
the luminosity, $\cal{L}$, is proportional to the
product of the mass depth of the target, $l$, and the number of
muon decays per second in the beam production straight section,
according to:
\begin{equation}
\cal{L}{\rm [cm^{-2}.s^{-1}]} = {\rm N_{Avo} \times
                           f_{ss} \times n_\mu\,[s^{-1}]} \times
                           \mathit{l} {\rm [g.cm^{-2}]},
\end{equation}
where ${\rm f_{ss}}$ is the fraction of the collider ring
circumference occupied by the production straight section,
$n_\mu$ is the rate at which each sign of muons is injected
into the collider ring (assuming they all circulate until
decay rather than being eventually extracted and dumped)
and the appropriate units are given in square brackets
in this equation and all later equations in this paper.
The proportionality constant is Avagadro's number,
${\rm N_{Avo}} = 6.022 \times 10^{23}$, since exactly one
neutrino per muon is emitted on average into the boosted
forward hemisphere, i.e. each muon decay produces two neutrinos
and half of them travel forwards in the muon rest frame.

 Because Avagadro's number is so large, the luminosities at
Mighty MURINEs will be enormous compared to those at collider
experiments. The luminosities for a reasonable scenario using
the workshop's straw-man parameter sets~\cite{hemc99intro} are
given in table~\ref{tab:EandL} along with, for
comparison, the final design goal luminosity for the HERA
ep collider. It can be seen that, roughly speaking, Mighty
MURINEs might achieve of order a million times the luminosity
of HERA. An ``accelerator year's'' running -- $10^7$ seconds --
at the 50 TeV MURINE's luminosity of
$2 \times 10^{37}\:{\rm cm^{-2}.s^{-1}}$
would amount to an impressive integrated luminosity of 200 inverse
{\em atto}barns per year while the even bigger straw-man luminosity at
5 TeV,
$1 \times 10^{39}\:{\rm cm^{-2}.s^{-1}}$, requires a luminosity
prefix that is even less familiar to the HEP community: 10 inverse
{\em zepto}barns per year.

\subsection{Center-of-Mass Energies for the Neutrino-Nucleon Interactions}
\label{subsec:expt_ECoM}

  It will be seen in section~\ref{sec:heavy} that HERA provides a useful
comparison for some of the physics capabilities of Mighty
MURINEs, particularly since the maximum center-of-mass
energies, ${\rm E_{CoM}}$, at the highest energy HEMCs might
even be comparable to those at the HERA collider. The electron-proton
${\rm E_{CoM}}$ at the collider is given by relativistic
kinematics as
\begin{equation}
{\rm E^{HERA}_{CoM} = 2 \sqrt{E_p E_e}},
\end{equation}
which is 314 GeV for the proton and
electron energies of the year 2000 upgrade to HERA,
${\rm E_p = 820}$ GeV and ${\rm E_e = 30}$ GeV.
For comparison, the MURINE's ${\rm E_{CoM}}$ is
\begin{equation}
{\rm E^{MURINE}_{CoM} = \sqrt{2 E_\nu M_pc^2 + (M_pc^2)^2}},
    \label{eq:MURINE_Ecom}
\end{equation}
where the proton mass corresponds to ${\rm M_pc^2} = 0.938$ GeV.
The neutrino energy can range right up to the muon beam energy,
${\rm E^{max}_\nu = E_\mu}$, and the energy spectrum
seen by the detector is relatively hard~\cite{numcbook}, with
an average neutrino energy within the $1/\gamma_\mu$ cone that
is 49\% of the muon beam energy. The
comparative center-of-mass energies for Mighty MURINEs and
HERA are summarized in table~\ref{tab:EandL}.

\begin{table}[ht!]
\caption{Energy, luminosity and event rates for Mighty MURINEs
at the 10 TeV and 100 TeV CoM muon colliders given in the HEMC'99
straw-man parameter
sets. The center-of-mass energy, ${\rm E_{CoM}}$,
is given for the neutrino-nucleon system.
The energy and luminosity at HERA are provided for
comparison. The event rates assume fractional straight section
lengths of ${\rm f_{ss}}= 0.02, 0.01$ for the 5+5 TeV and 50+50 TeV
parameter sets, respectively, and a detector mass-per-unit-area
of $l=1000\;{\rm g.cm^{-2}}$ that intercepts
the neutrino beam out to an angle $\theta_\mu = 1/\gamma_\mu$
subtended at the beam production
straight section. }
\begin{tabular}{|cccc|}
\hline
Facility & ${\rm E_{CoM}}$ & Luminosity, $\cal{L}$ & events/year \\
\hline
5 TeV MURINE        & 0 to 97 GeV  & $1 \times 10^{39}\:{\rm cm^{-2}.s^{-1}}$
                    & $1.7 \times 10^{11}$ \\
50 TeV MURINE       & 0 to 306 GeV & $2 \times 10^{37}\:{\rm cm^{-2}.s^{-1}}$
                    & $4 \times 10^{10}$ \\
HERA (2000 upgrade) & 314 GeV      & $7 \times 10^{31}\:{\rm cm^{-2}.s^{-1}}$
                    &  N.A.  \\
\end{tabular}
\label{tab:EandL}
\end{table}

\subsection{Cross Sections and Event Rates}
\label{subsec:expt_xsec}

 The event rate in the neutrino detector is a product of the
luminosity given in table~\ref{tab:EandL} and
the neutrino-nucleon scattering cross-section, which we now
discuss.

 The predominant interactions of 
neutrinos and anti-neutrinos at all energies above a few GeV
are charged current (CC) and neutral current (NC) deep inelastic
scattering (DIS) off nucleons ($N$, i.e. protons and neutrons)
with the production of several hadrons ($X$):
\begin{eqnarray}
\nu (\overline{\nu}) + N & \rightarrow & \nu (\overline{\nu}) + X
          \;\;\;\;\;\;\;(NC)
                                        \nonumber \\
\nu + N & \rightarrow & l^- + X 
          \;\;\;\;\;\;\;(\nu-CC)       
                                        \nonumber \\
\overline{\nu} + N & \rightarrow & l^+ + X
          \;\;\;\;\;\;\;(\overline{\nu}-CC),
                                        \label{eq:nuint}
\end{eqnarray}
where the charged lepton, $l$, is an
electron if the neutrino is an electron neutrino and a muon
for muon neutrinos.
The cross-sections for these processes are approximately
proportional to the neutrino
energy, $E_\nu$, with
numerical values of~\cite{quigg}:
\begin{equation}
 {\rm \sigma_{\nu N}\; for\;}
 \left(
 \begin{array}{c}
   \nu -CC \\
   \nu -NC \\
   \overline{\nu}-CC \\
   \overline{\nu}-NC
 \end{array}
 \right)\;
 \simeq
 \left(
 \begin{array}{c}
    0.72 \\ 0.23 \\ 0.38 \\ 0.13
   \end{array}
 \right)
\times 10^{-35}\: {\rm cm^2}
\times {\rm{E_\nu}[TeV]}
.
                                            \label{eq:xsec}
\end{equation}

  The number of events in the detector is easily seen to be
given by:
\begin{equation}
N_{events} =
             \cal{L} {\rm [cm^{-2}.s^{-1}] \times
             \rm 0.73 \times 10^{-35} \times
             0.49 \times E_\mu [TeV] \times T[s]},
\label{eq:Nevents}
\end{equation}
where T is the running time, $0.49 \times {\rm E}_\mu$ is
the average neutrino beam energy into the detector~\cite{numcbook}
and $0.73 \times 10^{-35}$ is the total cross-section-divided-by-energy
that is obtained from equation~\ref{eq:xsec} after summing over the
NC and CC interactions and averaging over neutrinos and anti-neutrinos.

  The final column of table~\ref{tab:EandL} shows the
impressive event sample sizes predicted from equation~\ref{eq:Nevents}:
up to of order $10^{11}$ events per year in a reasonably sized neutrino
target.

\subsection{High Performance Neutrino Detectors for Mighty MURINEs}
\label{subsec:expt_det}

\begin{figure}[t!] %
\centering
\includegraphics[height=3.5in,width=3.5in]{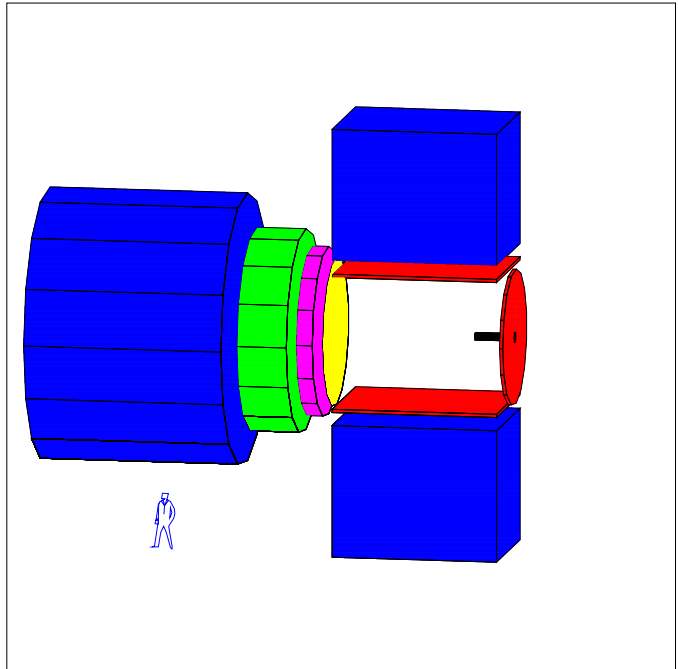}
\caption{Schematic example of a general purpose neutrino detector, reproduced
from reference~\cite{nufnal97}.
A human figure in
the lower left corner illustrates its size. The neutrino target is the small
horizontal cylinder at mid-height on the right hand side of the detector. Its
radial extent corresponds roughly to the radial spread of the neutrino pencil
beam, which is incident from the right hand side. Further details are given
in the text.}
\label{detector_fig}
\end{figure}

  The unprecedented event samples in small targets at MURINEs will
undoubtedly also spark a revolution in neutrino detector design and
performance, both to cope with event rates and to fully exploit the
physics potential of the beams. An example of a novel general
purpose neutrino detector that has been proposed
previously~\cite{nufnal97} for MURINEs is shown in
figure~\ref{detector_fig}.
The neutrino target is the cylinder at mid-height on the left hand
side of the figure. It comprises a stack of equally-spaced CCD
tracking planes, oriented perpendicular to the beam and with
spacings of order 1 mm, that provides
vertex tagging for events with hadrons containing charm or bottom
quarks.

 The general detector design of figure~\ref{detector_fig} should
remain appropriate for Mighty MURINEs although it would likely be
elongated to cope with the more boosted events, including perhaps
lengthening the target to several meters to increase the target
mass-per-unit-area and, correspondingly, the event rate.

  At these higher energies, the target mass-per-unit-area could be
increased still further by interspersing thin tungsten disks with
the CCD planes. There are two reasons why such a dense, high-Z target
becomes more practical than at lower energies: (i) multiple coulomb
scattering becomes less important at higher energies, so the tracking
resolution is degraded less, and (ii) the narrower pencil beam for Mighty
MURINEs allows the disks to have smaller radii than at lower energies --
smaller than the characteristic Moliere radius for electromagnetic
showers -- so it is speculated that electromagnetic showers will not
develop to excessively pollute the events, despite the large number
of radiation lengths along the axis of the target. (This assumption
needs to be checked in more detailed follow-up studies.)

 A specific scenario for the neutrino target that gives the 1000
${\rm g.cm^{-2}}$ target mass assumed in table~\ref{tab:EandL}
is as follows:
a 4 meter long target containing 4000 millimeter-long tracking
subunits, where each tracking subunit contains a thin tungsten disk of thickness
118 microns (0.227 ${\rm g.cm^{-2}}$) in front of a 100 micron thick
CCD pixel detector (0.023 ${\rm g.cm^{-2}}$). Each tungsten disk could
have a radius of 2 cm to match the beam radius at approximately 1 km
from production for a 5 TeV muon beam (as predicted from
equation~\ref{eq:thetanu}).
The CCD detectors can be wider than the beam radius to also track particles
moving outside the radial extent of the neutrino beam.

\begin{figure}[t!] %
\centering
\includegraphics[height=4.0in,width=5.0in]{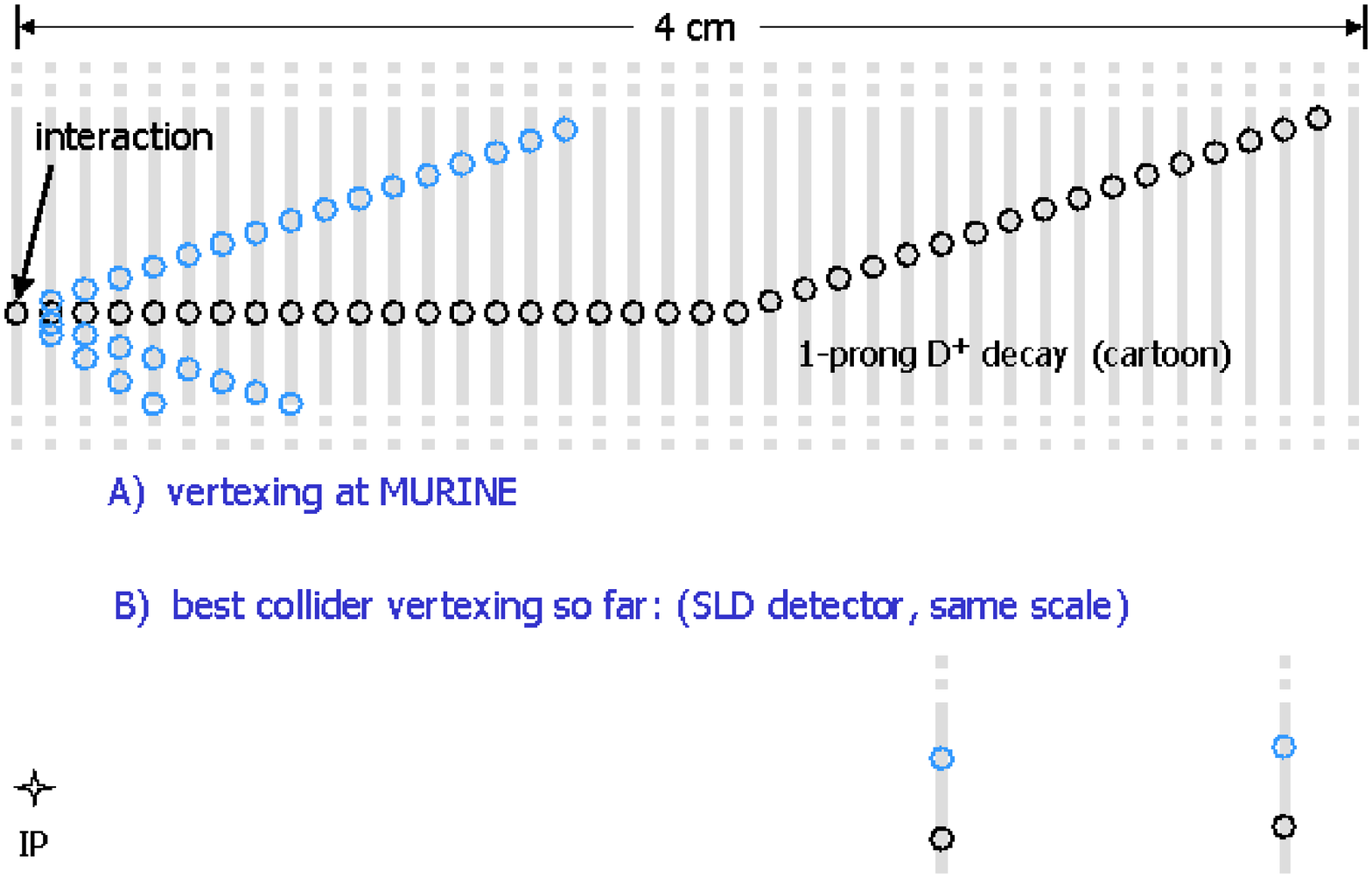}
\caption{
Conceptual illustration of the vertex tagging superiority
at MURINEs over that with collider experiment geometries. 
MURINEs could have a vertex plane of CCD pixel detectors
every millimeter. For comparison, the VXD3 vertexing detector
at the SLD~\cite{SLD}
experiment, which is universally regarded as the best existing
vertex detector in a collider experiment, has its two innermost
CCD tracking planes at 2.8 cm and 3.8 cm from the interaction point.
A cartoon of a 1-prong ${\rm D^+}$ decay has been drawn to
illustrate the advantages of closely spaced vertex detectors.
For clarity of illustration, the kink deflection angle has
been drawn much larger than would be typical. The 2 cm distance
to decay for the ${\rm D^+}$ charmed meson corresponds to
the average boosted lifetime for a 120 GeV  ${\rm D^+}$.
Most charm and beauty hadrons at a Mighty MURINE, having
much higher energies than this, will travel even further
and hence traverse even more planes of tracking before decaying.
}
\label{vertexing}
\end{figure}

 The vertex tagging performance of the target in figure~\ref{detector_fig}
is expected~\cite{nufnal97} to be better than any other existing or planned
high-rate detector for heavy quark physics, and should continue
to improve with beam energy. Mighty MURINEs
should attain close to 100 percent efficiency for both c and b (easier)
vertex tagging in the target (excepting all-neutral decay modes, of course)
since the average boosted lifetime for TeV-scale charm and beauty hadrons --
of order 10 cm -- would span many planes of CCD's. The extremely favorable
geometry for vertexing is illustrated in figure~\ref{vertexing},
where it is also compared
with the vertexing geometry at a collider detector.

 As with lower energy MURINEs, the detector backing the neutrino target
should faithfully reconstruct both CC and NC
event kinematics. The lower energy MURINEs provide essentially full
particle identification through the muon toroids and dE/dx plus
cherenkov radiation in the TPC. The particle ID for long-lived charged
hadrons
would become more difficult at higher energies although, speculatively,
the cherenkov radiation in an elongated TPC might still give effective
PID for particle energies up to a couple of hundred GeV -- this requires
further study.
To somewhat compensate, the trajectories of most photons should be
very well measured when they convert in the stack of tungsten disks
that comprise most of the target mass. This will be particularly helpful
for the reconstruction of neutral pions.

\section{MURINEs: A New Realm for Neutrino Physics}
\label{sec:MURINE}

  This subsection gives a brief non-technical overview of the
high rate neutrino physics topics expected for MURINEs in general.
Topics that might be expanded on with the higher energies of Mighty
MURINEs are pointed out in preparation for more detailed discussion
in the following sections.

\subsection{Neutrino Interactions with Quarks}
\label{subsec:MURINE_phys}

  Put simply, the DIS interactions of equation~\ref{eq:nuint}
involve a simple projectile
(the neutrino),
interesting interactions (the CC and NC weak interactions) and
a complicated target (the nucleon). The bulk of the
physics interest lies in the interactions with the quark
constituents rather than in the properties of the
neutrinos themselves.
The complementary analyses that study the potential
oscillations of neutrino flavors tend to
be more the domain of lower energy MURINEs, at muon energies
of order 100 GeV or below, and won't be discussed further here.

 By the TeV-energy scale and above, the CC and NC interactions of
equation~\ref{eq:nuint} have become very well described as being
the quasi-elastic
(elastic) scattering of neutrinos off one of the many quarks
(and anti-quarks), q, inside the
nucleon:
\begin{eqnarray}
\nu (\overline{\nu}) + q & \rightarrow & \nu (\overline{\nu}) + q
          \;\;\;\;\;\;\;\;\;(NC)
                                        \label{eq:ncnuq} \\
\nu + q^{(-)} & \rightarrow & l^- + q^{(+)} 
          \;\;\;\;\;\;\;(\nu-CC)       
                                        \label{eq:ccnuq} \\
\overline{\nu} + q^{(+)} & \rightarrow & l^+ + q^{(-)}
          \;\;\;\;\;\;\;(\overline{\nu}-CC).
                                        \label{eq:ccnubarq}
\end{eqnarray}
The CC and NC interactions are mediated through the exchange of a
virtual W or Z boson, respectively.
All quarks -- up quarks (u), down quarks (d) and the smaller
``seas'' of the progressively heavier strange (s), charm (c)
and even beauty (b) quarks -- participate in NC scattering interactions
of both neutrinos and anti-neutrinos. In contrast, charge conservation
specifies the charge sign of the quarks participating in the
CC processes as indicated by the labels:
$q^{(-)} \in d,s,b,\overline{u},\overline{c}$ and
$q^{(+)} \in u,c,\overline{d},\overline{s},\overline{b}$.
The hadrons seen in the detector are produced by the
``hadronization'' of the final state struck quark at the nuclear
distance scale.

\subsection{The Intrinsic Richness of Neutrino Physics}
\label{subsec:MURINE_rich}

 Experimentally,
the interaction type of almost all events can be distinguished 
with little ambiguity by the charge of the final
state lepton (and its flavor: i.e. electron
or muon): neutral (an unseen neutrino), negative or
positive for the 3 respective processes in equation~\ref{eq:nuint}.

 It is seen that equations~\ref{eq:ncnuq} through~\ref{eq:ccnubarq}
probe 3 different weightings of the quark flavors inside a nucleon,
through weak interactions involving both the W and Z. For comparison,
only a single and complementary weighting is probed by the best
competitive process -- the photon exchange interactions of
{\em charged} lepton scattering experiments at HERA and fixed
target facilities. Much of the uniqueness and richness of neutrino
scattering physics derives from this variety of interaction
processes.

\subsection{Physics Topics at MURINEs}
\label{subsec:MURINE_topics}

 Mighty MURINEs will extend and improve on the already
considerable range of unique topics that will have been
explored at lower energy MURINEs. The interested reader
is referred to reference~\cite{numcbook} for more details.
Here we list those topics that will have already been well
addressed in earlier MURINEs and comment on any added potential
that might be available using the higher energies at Mighty MURINEs.


\subsubsection{Probing Nucleon Structure}

 The redundant probes of the proton's and neutron's internal
structure should provide some of the most precise measurements and tests
of perturbative QCD - the theory of the strong interaction - and
will also be invaluable input for many analyses at pp colliders.
Neutrinos are also intrinsically 100\% longitudinally polarized,
so experiments with polarized targets could additionally map out
the spin structure of the nucleon. Some of these analyses might
obtain modest benefits from the higher energies and statistics
at Mighty MURINEs.

\subsubsection{Precision Electroweak Measurements}

 Besides using W and Z exchange as nuclear probes, the
interactions themselves provide important precision tests
of the standard model of elementary particles. Two
measurements of total interaction cross sections will provide
determinations of the fundamental weak mixing angle parameter
of the electroweak theory, $\sin^2 \theta_W$, from
(i) the ratio of NC to CC total cross sections and (ii) the
absolute cross section for the rarer process of neutrino-electron
scattering, which is 3 orders of magnitude less common than
neutrino-nucleon scattering. In both cases, the fractional
uncertainties in $\stw$ might approach the $10^{-4}$ level,
which would be complementary and competitive to the best
related measurements in collider experiments.

  The first of the two measurements will already be systematically
limited at MURINEs~\cite{numcbook} so large gains should not be
expected at Mighty MURINEs. The situation is not so clear for
the electron scattering process, where the higher event statistics
could still be beneficial. Speculatively, these higher statistics might
also allow the use of liquid hydrogen targets with improved
experimental capabilities.

\subsubsection{Charm and Beauty Factories}

  MURINEs of all energies will be excellent charm factories, with
of order 1\% to 10\% of the events containing a charmed hadron, depending
on the beam energy. TeV-scale MURINEs and above will also produce enough
B hadrons to be considered as beauty factories and Mighty MURINES might
be very impressive B factories, as will be discussed in section~\ref{sec:B}.

\subsubsection{Quark Mixing Studies}

 There is much additional interest in experimentally
partitioning the CC event sample to obtain
the partial cross sections for the various possible quark flavor
transition combinations represented by the 
$q^{(-)}$ and $q^{(+)}$ symbols in equations~\ref{eq:ccnuq}
and~\ref{eq:ccnubarq}. MURINEs, in general, should have the quark-tagging
capability to separate the various quark flavor contributions,
as was discussed in the preceding
section. Mighty MURINES, with their extra capability
for producing heavy final-state quarks, could make great
strides beyond previous MURINEs for these ``quark mixing''
studies, as will be expanded on in section~\ref{sec:qm}.

\subsubsection{Rare and Exotic Processes}

   The higher statistics and, particularly, energies available at 
Mighty MURINEs would clearly expand the scope for studies of
rare processes and searches for exotic processes. This will
be covered in section~\ref{sec:heavy}.

\section{Mighty MURINEs as B Factories}
\label{sec:B}

  The charged current production of b quarks off the light quarks
in the nucleon is heavily suppressed due to small off-diagonal CKM
matrix elements. However, the fraction of neutrino-induced events
containing B hadrons rises rapidly with energy~\cite{numcbook} due to
the decreasing threshold suppression for two higher-order processes
involving gluons in the initial state:
\begin{enumerate}
  \item  $\bbbar$ pair production in neutral current interactions:
\begin{equation}
   \nu N \rightarrow \nu b \overline{b} X.
       \label{bbprod}
\end{equation}
  \item  charged current production of $\cbbar$ and $\bcbar$ from the
      charged current interactions of neutrinos or anti-neutrinos:
 \begin{equation}
   \nu N \rightarrow l^- \overline{b} c X
       \label{bbprodnu}
\end{equation}
and
\begin{equation}
  \overline{\nu} N \rightarrow l^+ b \overline{c} X,
       \label{bbprodnubar}
\end{equation}
respectively.
\end{enumerate}

  Preliminary estimates~\cite{Timcomm, numcbook} for the fraction of events
from each of these processes are tabulated versus neutrino energy in
table~\ref{tab:bprod}. The second of the two processes is seen
to be less common than the first. To compensate, it provides an
extremely pure and efficient tag to distinguish between b and
anti-b quark production: b production is always accompanied by a
positive primary lepton (from anti-neutrino interactions) and
anti-b production by a negative primary lepton (from neutrino interactions).
This will be very helpful for studies of oscillations of
${\rm B_0}$'s and ${\rm B_S}$'s.

\begin{table}[ht!]
\caption{Fraction of events~\cite{Timcomm} producing B's in the
final states $\bbbar$, $\bcbar$ or $\cbbar$, from neutrinos
and anti-neutrinos of energies 1 TeV and 10 TeV. Estimates are
preliminary.}
\begin{tabular}{|cccc|}
\hline
$\nu$ or $\nubar$  & $\Enu$ & final state  & fraction \\
\hline
$\nu$      &  1 TeV   & $\bbbar$  & $6 \times 10^{-4}$  \\
$\nubar$   &  1 TeV   & $\bbbar$  & $6 \times 10^{-4}$  \\
$\nu$      &  1 TeV   & $\cbbar$  & $2 \times 10^{-5}$  \\
$\nubar$   &  1 TeV   & $\bcbar$  & $2 \times 10^{-5}$  \\
\hline
$\nu$      &  10 TeV  & $\bbbar$  & $4 \times 10^{-3}$  \\
$\nubar$   &  10 TeV  & $\bbbar$  & $4 \times 10^{-3}$  \\
$\nu$      &  10 TeV  & $\cbbar$  & $8 \times 10^{-5}$  \\
$\nubar$   &  10 TeV  & $\bcbar$  & $6 \times 10^{-5}$  \\
\end{tabular}
\label{tab:bprod}
\end{table}

  Combining the numbers in tables~\ref{tab:EandL} and~\ref{tab:bprod}
predicts
event rates of perhaps $10^8$ to $10^9$ B's per year at Mighty MURINEs.
This is intermediate between the expectations of the $\ee$ B
factory experiments ($\sim 10^7$ events/year) and the hadron
B factories, HERA-B, BTeV and LHC-B (up to $\sim 10^{11}$ events/year,
with up to a few times $10^9$ events tagged for analysis).
As already mentioned, however,the vertexing
capabilities and other experimental conditions at Mighty MURINES should
be superior in some aspects to those at the $\ee$ B factories and vastly
superior to the very difficult experimental conditions at the hadronic
B factories.

 Three speculative examples of B analyses that would benefit from the
unique experimental conditions at Mighty MURINEs are:
\begin{itemize}
   \item   the superior vertexing capabilities should be ideal for studying
the expected fast oscillations of ${\rm B_s}$'s, perhaps following up on
previous B factories with more precise measurements of the oscillation
frequency and greater sensitivity to any asymmetry in the
${\rm B_s}$ and ${\rm \overline{B_s}}$ decay rates
   \item    some studies of the B baryons, $\Lambda_b$,
$\Xi_b^-$ and $\Xi_b^0$, which are not produced in $e^+e^-$ B factories,
may also plausibly be best performed at a Mighty MURINE
   \item   it might have a chance~\cite{Bigi} to measure the branching
ratio for the all-neutral rare decay $B_d \rightarrow \pi^0 \pi^0$,
which is expected to be of order $10^{-6}$.
This would provide an estimate for the otherwise problematic
``penguin-diagram pollution'' in the analogous charged pion decay
$B_d \rightarrow \pi^+ \pi^-$, and this could go some way to resurrecting
the charged decay mode as one of the central CKM processes at B factories.
However, observing the neutral decay mode does not look feasible at any
future B factories other than Mighty MURINEs.
\end{itemize}

   The final process deserves some further explanation since the decay
itself doesn't provide a vertex. However, the close to 100\% vertex
reconstruction efficiency could instead act as a veto to reduce the
backgrounds from the pair-produced B's in neutral current interactions.
The signature for the signal process would be a neutral current event with
(i) a single vertex from the other B, (ii) 4 converted gammas reconstructing
to 2 high energy $\pi^0$'s that, in turn,
reconstruct to the $B_d$ mass and (iii) no
suspicion of another B or charm vertex. Hence, the analysis -- although
admittedly still exceedingly difficult -- would benefit from both the
exceptional vertexing and neutral pion reconstruction at Mighty MURINEs.

  Therefore, to summarize this section, the initial expectation is that
Mighty MURINEs should be able to do follow-up precision studies in at least
some of the most difficult areas of B physics, even after the other B
factories have run.

\section{Quark Mixing: Measurements Beyond the B Factories}
\label{sec:qm}

 This section enlarges on the theoretical interest in measurements
of quark mixing at MURINEs and also provides detail on the central
role that Mighty MURINEs could assume if they reached sufficient
energies to begin producing top quarks.

\subsection{Theoretical Interest in the CKM Matrix}
\label{subsec:qm_theory}

 Quark mixing is one of the least understood and most
intriguing parts of elementary particle physics, and the
confinement of quarks inside hadrons also makes it one of
the hardest areas to study. The CC weak interaction for quarks
differs from this interaction for leptons by mixing quarks from different
families, i.e. any positively charged quark,
$q^{(+)} \in u,c,t$, has some probability of
being converted into any of its negatively charged
counterparts, $q^{(-)} \in d,s,b$, and vice versa,
rather than being uniquely associated with its
same-family counterpart (i.e. $d \leftrightarrow u$,
$s \leftrightarrow c$ and $b \leftrightarrow t$).
This feature is accommodated in the standard model of
elementary particle physics through the unitary 3-by-3
Cabbibo-Kobayashi-Maskawa (CKM) matrix,
${\rm V}_{ij}$, that connects the positively charged
$q^{(+)}_i$'s with the three $q^{(-)}_j$'s.
(The corresponding matrix for leptons is trivially
the 3-by-3 identity matrix.)

 Apart from verifying that the SM description for quark mixing
is indeed correct, the CKM matrix has additional interest through
its hypothesized association with CP violation: the puzzling
phenomenon that some particle properties, such as decay rates,
have been found to differ slightly from those of the corresponding
anti-particles. CP violation could also have cosmological
implications; it has been invoked as one possible explanation
for the comparative scarcity of anti-matter in the universe.
The presence of a complex phase in the CKM matrix is the largely
{\em untested} standard model explanation/parameterization for
CP violation.

 It is a testament to the perceived importance of the CKM matrix
and CP violation that much of today's experimental HEP effort is
devoted such studies, including B and phi factory colliders,
LHC-B, HERA-B, B-TeV, K-TeV and many others.
Neutrino-nucleon scattering has impressive potential to
augment these studies but, until the arrival of MURINEs, it
will be held back by inadequate beam intensities.

\subsection{Quark Mixing Studies at MURINEs}
\label{subsec:qm_MURINE}

\begin{figure}[t!] %
\centering
\includegraphics[height=2.5in,width=3.5in]{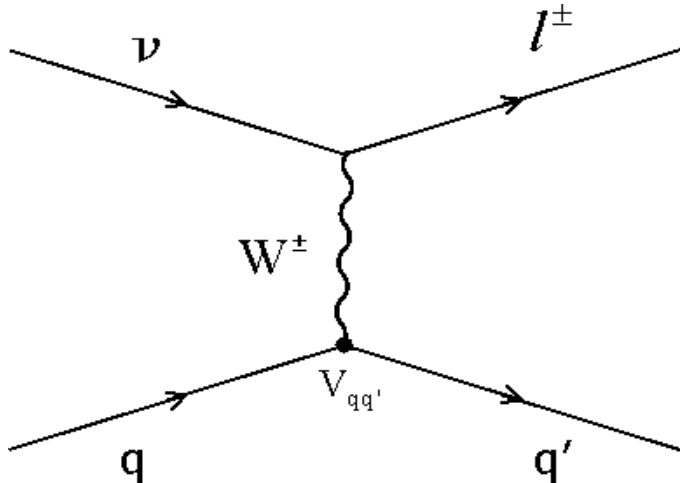}
\caption{
The Feynman diagram for charged current neutrino-quark scattering,
showing that the CKM matrix element $V_{qq'}$ participates as
an amplitude in the W-quark vertex.
}
\label{Vqqdiagram}
\end{figure}

  The new neutrino studies at MURINEs will be much cleaner
theoretically than most of the other experimental processes
and will offer much complementary information.

  Figure~\ref{Vqqdiagram}
is the Feynman diagram for the basic scattering process,
showing that the CKM matrix element $V_{qq'}$ participates as
an amplitude in the W-quark coupling. As a fundamental difference
between $\nu$N DIS and all other types of CKM measurements, the
scattering process involves the interaction of an external
W boson probing the quarks inside a nucleon rather than an internal
W interaction inside a hadron that, e.g., initiates a B decay.
(In principle, the HERA ep collider could also do measurements
involving an external W exchange, but these turn out not to be
feasible in practice~\cite{HERA_CKM}.) This is a substantial
theoretical advantage for neutrino scattering because the
``asymptotic freedom'' property of QCD predicts quasi-free quarks
with reduced influence from their hadronic environment at the
higher 4-momentum-transfer (Q) scales available with an external
W exchange.

  The CKM measurements at MURINEs will be complementary
to, say, the CKM measurements at B factories in that the measurements
are of the magnitudes of individual CKM matrix elements rather than
of interference terms involving pairs of elements. As can be inferred
from figure~\ref{Vqqdiagram}, this arises because the differential
cross-sections, $\frac{d\sigma}{dx}(q_i \rightarrow q_j)$,
for the quark transitions are proportional
to the absolute squares of the CKM elements:
\begin{eqnarray}
\frac{d\sigma}{dx}( d\rightarrow c) \propto x\, d(x)|V_{cd}|^2
                    \times T(m_c, x)
\label{eq:Vcd} \\
\frac{d\sigma}{dx}( s\rightarrow c) \propto x\,s(x)|V_{cs}|^2
                    \times T(m_c, x)
\label{eq:Vcs} \\
\frac{d\sigma}{dx}( u\rightarrow b) \propto x\,u(x)|V_{ub}|^2
                    \times T(m_b, x)
\label{eq:Vub} \\
\frac{d\sigma}{dx}( c\rightarrow b) \propto x\,c(x)|V_{cb}|^2
                    \times T(m_b, x)
\label{eq:Vcb} \\
\frac{d\sigma}{dx}( d\rightarrow t) \propto x\,d(x)|V_{td}|^2
                    \times T(m_t, x)
\label{eq:Vtd} \\
\frac{d\sigma}{dx}( s\rightarrow t) \propto x\,s(x)|V_{ts}|^2
                    \times T(m_t, x)
\label{eq:Vts} \\
\frac{d\sigma}{dx}( b\rightarrow t) \propto x\,b(x)|V_{tb}|^2
                    \times T(m_t, x),
\label{eq:Vtb}
\end{eqnarray}
where the Bjorken scaling
variable, $x$, with $0<x<1$, is a relativistically invariant
quantity that can be reconstructed for each event and, roughly
speaking, measures the fraction of the nucleon's 4-momentum
carried by the struck quark. The respective
initial-state quark densities as functions of Bjorken $x$ have been
labeled $d(x)$, $s(x)$, $u(x)$, $c(x)$ and $b(x)$, and the
$T(m_q, x)$'s are threshold suppression
factors due to the masses, $m_q$, of the final-state quarks.

 The $T(m_q, x)$ mass suppression factors are zero or much less
than unity for all $x$ below neutrino energies that
can readily supply enough CoM energy to produce the massive
final state quarks. From equation~\ref{eq:MURINE_Ecom}, the
$T(m_q, x)$'s will asymptotically approach unity only for muon beam
energies such that: 
\begin{equation}
m_q^2 \ll  2 M_p E_\mu/c^2 + M_p^2.
\end{equation}
This places the following lower bounds on beam energies for the
efficient production of charm, beauty and top quarks, respectively:
\begin{eqnarray}
{\rm m_c \sim 1.3-1.7\:GeV/c^2\;\;} &
     \Rightarrow {\rm \;\;E_\mu \gg 1\;GeV } \\
      \nonumber
{\rm m_b \sim 5\:GeV/c^2\;\;} &
     \Rightarrow {\rm \;\;E_\mu \gg 13\;GeV } \\
      \nonumber
{\rm m_t \sim 175\:GeV/c^2\;\;} &
     \Rightarrow {\rm \;\;E_\mu \gg 16\;TeV }.
      \label{eq:tthres}
\end{eqnarray}

  The extraction of the CKM matrix elements from the MURINEs'
experimental data will be analogous to, but vastly superior to,
current neutrino measurements of $\Vcd$, the only CKM matrix
element that is currently best measured in neutrino-nucleon
scattering~\cite{Vcdmeas}.
 The experimentally determined event counts and kinematic
distributions of the
quark-tagged event samples provide measurements of the differential
distributions for each of the final state quarks.
The differential cross-sections, 
$\frac{d\sigma}{dx}(q_i \rightarrow q_j)$,
and CKM matrix elements, $\Vij$, are derived from
equations~\ref{eq:Vcd} through~\ref{eq:Vtb} using
some auxiliary knowledge of the quark x-distributions within the nucleons
and also a model for the mass threshold suppression terms,
$T(m_q, x)$. In practice, this information should be obtainable
largely from the data samples themselves: from CC and NC structure
function measurements and the observed kinematic dependences in
the heavy quark event sample.

\subsection{Expected Measurement Precisions at MURINEs}
\label{subsec:qm_precis}

 Today's measurements of $\Vcd$ in $\nu$N scattering are already
the most precise in any process, despite the coarse instrumentation
of the neutrino detectors and the consequent requirement to use
the semi-muonic subsample of charm decays for final
state charm tagging. Even the lowest energy MURINEs under consideration
(dedicated neutrino factories with $E_\mu \simeq 10$ GeV and up)
will provide an opportunity~\cite{numcbook} to extend
to unique and precise measurements of the elements $\Vcd$ and
probably $\Vcs$, now using vertex tagging of charm and with
much improved knowledge of the quark distributions.

  Further measurements of the more theoretically interesting
elements $\Vub$ and $\Vcb$ will become available at MURINES
with muon energies of around 100 GeV or above, which can
provide high enough neutrino energies for B production.
The B-production analyses at these higher energy MURINEs should be
experimentally rather similar to the charm analyses but would
have vastly greater theoretical interest.

   Both $\Vub$ and $\Vcb$
determine the lengths of sides of the ``unitarity triangle''
that is predicted to exist if the CKM matrix is indeed
unitary~\cite{PDG_CKM}.
The main goal of today's B factories is to measure the
interior angles of this triangle to confirm that it is
indeed a triangle, and the complementary input from a MURINE
will be an enormous help in this verification process. In
particular, the predicted~\cite{nufnal97,numcbook} 1-2 \%
accuracy in $\Vub^2$ is several times better than predicted
accuracies in any future measurements of other processes, and will
obviously provide a very strong constraint on the unitarity
triangle.

  Predicted experimental accuracies for a 500 GeV
MURINE are summarized in table~\ref{tab:ckm}. These
measurements would likely be improved still further with
the higher event statistics and cleaner theoretical
analysis available at a Mighty MURINE.

\begin{table}[ht!]
\caption{
Predicted CKM measurements for lower energy MURINES,
reproduced from reference~\cite{jh99}. The first row
for each element, in bold face, is the absolute square
of that matrix element, which is proportional to the
experimental event rate --
see equations~\ref{eq:Vcd} through~\ref{eq:Vtb}.
The second row for each element gives
current percentage uncertainties in the absolute squares,
without applying unitarity constraints,
and speculative projections of the uncertainties after analyses
from a 500 GeV MURINE. The measurements of $\Vcd^2$ and $\Vcs^2$
might be comparably good even for a 50 GeV MURINE but
 $\Vub^2$ and $\Vcb^2$ would not be measured.
}
\begin{tabular}{|c|lll|}
\hline
          & \hspace{0.2 cm} \bf{d} & \hspace{0.2 cm} \bf{s}  &
                                          \hspace{0.3 cm}\bf{b}  \\
\hline
\bf{u}    &   \bf{0.95}  &  \bf{0.05}    &  \bf{0.00001}  \\
          &   $\pm$0.1\%  &  $\pm$1.6\%    & $\pm$50\% $\rightarrow$ 1-2\% \\
          &&& \\
\bf{c}    &   \bf{0.05}  &  \bf{0.95}    &  \bf{0.002}  \\
          &   $\pm$15\% $\rightarrow$ 0.2-0.5\%   &
              $\pm$35\% $\rightarrow$ $\sim 1$\%         &
              $\pm$15\% $\rightarrow$ 3-5\%     \\
          &&& \\
\bf{t}    &  \bf{0.0001}  &  \bf{0.001}  &  \bf{1.0} \\
          &   $\pm$25\%   &  $\pm$40\%   & $\pm$30\%
\label{tab:ckm}
\end{tabular}
\end{table}

\subsection{Possible Measurements of ${\rm V_{td}}$ in Flavor
Changing Neutral Current Interactions}
\label{subsec:qm_Vtd}

\begin{figure}[t!] %
\centering
\includegraphics[height=2.5in,width=3.5in]{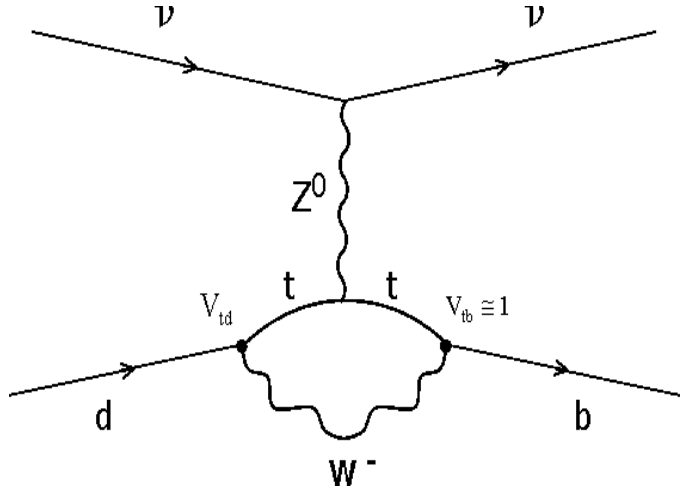}
\caption{
The Feynman diagram for B production through a flavor changing
neutral current interaction involving a top quark loop.
}
\label{Vdbdiagram}
\end{figure}

\begin{figure}[t!] %
\centering
\includegraphics[height=2.5in,width=3.5in]{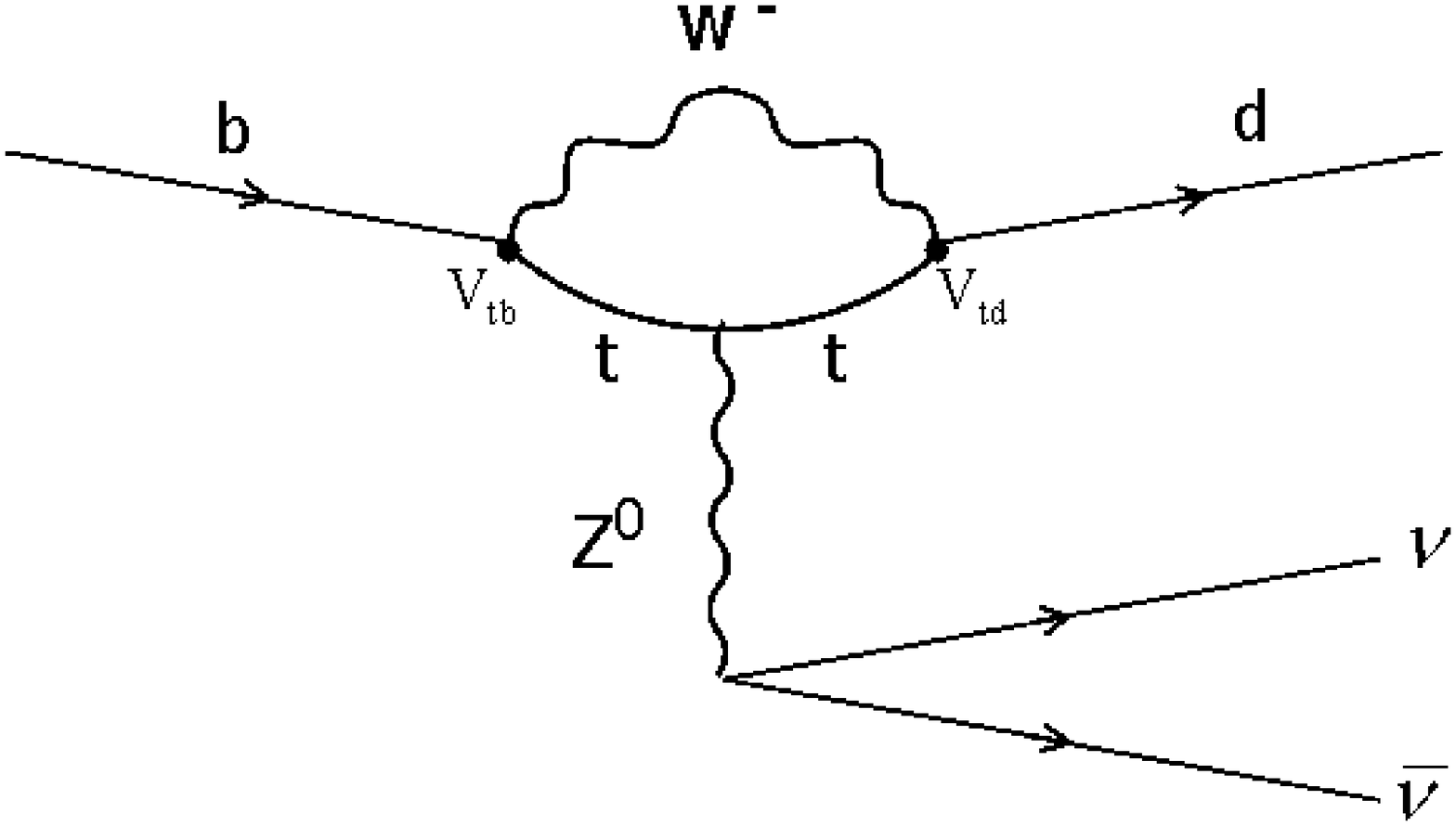}
\caption{
The Feynman diagram for the rare B decay
${\rm B \rightarrow X_d \nu \overline{\nu} }$, which
is analogous to the neutrino process of figure~\ref{Vdbdiagram}.
}
\label{b_to_dnunu}
\end{figure}

 The increased event statistics and neutrino energies at
Mighty MURINEs might even allow the measurement of the further
matrix element $\Vtd$ through the flavor changing neutral
current (FCNC) interaction of figure~\ref{Vdbdiagram}.

 This process is analogous to the predicted rare B decay
${\rm B \rightarrow X_d \nu \overline{\nu} }$, shown in
figure~\ref{b_to_dnunu}, where ${\rm X_d}$ represents
inclusive production of hadrons containing a down quark.
As related measurements, the very rare kaon decay processes
$K^- \rightarrow \pi^- \nu \overline{\nu}$
and $K^0_L \rightarrow \pi^0 \nu \overline{\nu}$
proceed through diagrams equivalent to~\ref{b_to_dnunu}
except with the incoming b quark replaced by an s quark
and, correspondingly, ${\rm V_{tb}}$ replaced by
${\rm V_{ts}}$. Therefore, the charged kaon decay
has the potential to measure $|{\rm V^*_{ts} V_{td} }|$
and its neutral counterpart actually measures
the imaginary part of this quantity,  ${\rm Im (V^*_{ts} V_{td}) }$,
due to ${\rm K^0-\overline{K^0}}$ interference. One event
of the decay $K^- \rightarrow \pi^- \nu \overline{\nu}$
has been seen so far~\cite{E787}, consistent with its predicted
tiny branching ratio of $(8.2 \pm 3.2) \times 10^{-11}$,
and the even rarer neutral decay process has yet to be
observed.

  The search for the B decay signal looks to be extremely
challenging even at future B factories, so it is unlikely
to yield an accurate measurement of $\Vtd$. Therefore, a
neutrino measurement of this quantity, at or below the 10\% level,
might still be valuable even after the B factories have run,
augmenting the complementary measurements, perhaps eventually
with comparable accuracy~\cite{Littenberg}, of
$|{\rm V^*_{ts} V_{td} }|$ and ${\rm Im (V^*_{ts} V_{td}) }$
expected in future generations of rare kaon decay experiments
and the precise measurement of the ratio $\Vts/\Vtd$
that is to be eventually expected from ${\rm B_d}$ and ${\rm B_s}$
oscillations.

 For neutrino energies well above the B production threshold,
the process of figure~\ref{Vdbdiagram} will occur at the
level~\cite{numcbook} of order $10^{-8}$ of the total neutrino-induced
event sample unless some exotic physics process intervenes to
increase the production rate. The signature is production
of single ${\rm B^-}$ mesons from the valence d quarks at high
x, whose rate should be directly proportional to the product
of $\Vtd$ and the known valence quark density in nuclons.

 The main background will come from b--anti-b production
where the partner B meson containing the b anti-quark
has escaped detection from either its primary decay vertex
or through the decay vertex of its daughter charmed meson.
The background process is easily separable from the signal
on a statistical basis because it is symmetric in
${\rm B^-}$ versus ${\rm B^+}$ mesons. However, the raw
production rate~\cite{numcbook} is roughly five orders of
magnitude above the signal so the statistical viability
of the analysis would require the raw background to be
reduced by perhaps 4 to 5 orders of magnitude.
This can be contemplated
only because of (1) the very different event kinematics, with
almost all of the background events at low Bjorken $x$ and the signal
mostly at high $x$, and (2) the unprecedented veto power for B
decays that is expected in the vertexing detectors at MURINEs.
Even so, a raw signal event sample of thousands of events might
be needed for a 10\% measurement of $\Vtd$ given the statistical
dilution that background processes might entail. This would
require several years running for the experimental parameters
of table~\ref{tab:EandL}.

\subsection{CKM Measurements from the Production of Top Quarks at
the Highest Energy Mighty MURINEs}
\label{subsec:qm_top}

 Aside from top production in loops,
a daunting leap of 3 orders of magnitude in beam energy
would be required to move from the CKM elements involving
B production to those involving top production, as is
seen from comparing the second and third rows of equation~\ref{eq:tthres}.
 Uniquely precise direct measurements of ${\rm |V_{td}|^2}$ and
${\rm |V_{ts}|^2}$ and, possibly, ${\rm |V_{tb}|^2}$
from the production of top quarks will become available if and
when muon colliders eventually reach the 100 TeV CoM energy scale.
(Note that muon collider energies even up to 1000 TeV, i.e. 1 PeV,
have been speculated, using muon acceleration in linacs~\cite{Zimmermann}.)

  Such impressive machines are prospects for the
far distant future, and would be intended to zero in on
a coherent understanding of the elementary building blocks
of our universe. It should be stated that a major sea change
from current theoretical prejudices would be required if the
CKM matrix and its information on CP violation was to become
central to the construction or verification of such a ``theory
of everything''. Disregarding the current prejudices,
top production at these highest energy Mighty MURINEs
would move the experimental probing of the CKM matrix to a
level of accuracy that appears to be inaccessible to any other
type of experiment.

  As will be explained further in the following section, a simple
scaling from top production estimates calculated for
HERA~\cite{HERAref} predicts of order $10^5$ top quark events
for 1 inverse zeptobarn of integrated luminosity at muon energies
slightly above 50 TeV. (A more accurate and detailed calculation
is obviously required!)

 Experimentally, top production should be relatively easy
to tag with very high efficiency and purity. The two signatures are:
\begin{eqnarray}
\nu_\mu N \rightarrow \mu^- {\rm (2\;jets)} {\rm (b\; jet)}
    \label{eq:qm_topjetsig} \\
\nu_\mu N \rightarrow \mu^- l^+ \nu {\rm (b\; jet)},
    \label{eq:qm_toplepsig}
\end{eqnarray}
with 68\% and 32\% BR's, respectively.
Because of the large top mass, the final state jets can each have
large acoplanarities, and the rarity of backgrounds with b quarks
makes both signatures particularly distinctive. Additionally,
in the first case the 2 other jets will reconstruct to the W
mass while the presence of a second high-$\pt$, high energy lepton
and large missing $\pt$ from the neutrino will make the second signature
even more striking.

  No attempt will be made to even guess at the measurement accuracy.
As a general comment, the beam energy will never be very far above
the threshold for top quark production, so
the feasibility and accuracy of the measurements would depend
more strongly on the muon beam energy than the beam intensity.
In almost all cases,
the measurements of the CKM matrix elements involving top should
be statistically limited because of the relatively small statistics
(except at PeV-scale colliders!), their distinctive experimental
signature and the accurately predictable threshold behavior.
The sequentially decreasing populations at high $x$ of the
progressively heavier initial state quarks -- d, s and b --
should compensate or over-compensate the trend for the higher
couplings to the top quark in the respective measurements of
$\Vtd$, $\Vts$ and $\Vtb$. Whether the first, the first two
or all three matrix could be measured would presumably
also depend strongly on the beam energy.

 CKM measurements involving top would extend CKM studies beyond
the paradigm
of the unitarity triangle that is connected to B factory studies.
For example, the conventional unitarity triangle is formed from
the dot product of columns 1 and 3 of the CKM matrix~\cite{PDG_CKM}.
Measurements of the CKM elements involving the top quark
would also provide enough experimental input to test the corresponding
triangle involving {\em rows} 1 and 3 of the matrix, which is of
comparable theoretical value in exploring CP violation. The analysis
of experimental results would more likely be couched in more general
theoretical terms, including unitarity tests and global fits to
the 4 parameters -- three magnitudes and a phase -- that characterize
the unitary 3-by-3 CKM matrix. Consistency of these fits would
probe the SM hypothesis at a level that would not be possible
without Mighty MURINEs.

\section{Heavy Particle Production -- Mighty MURINEs Versus HERA}
\label{sec:heavy}

\begin{figure}[t!] %
\centering
\includegraphics[height=3.5in,width=6.0in]{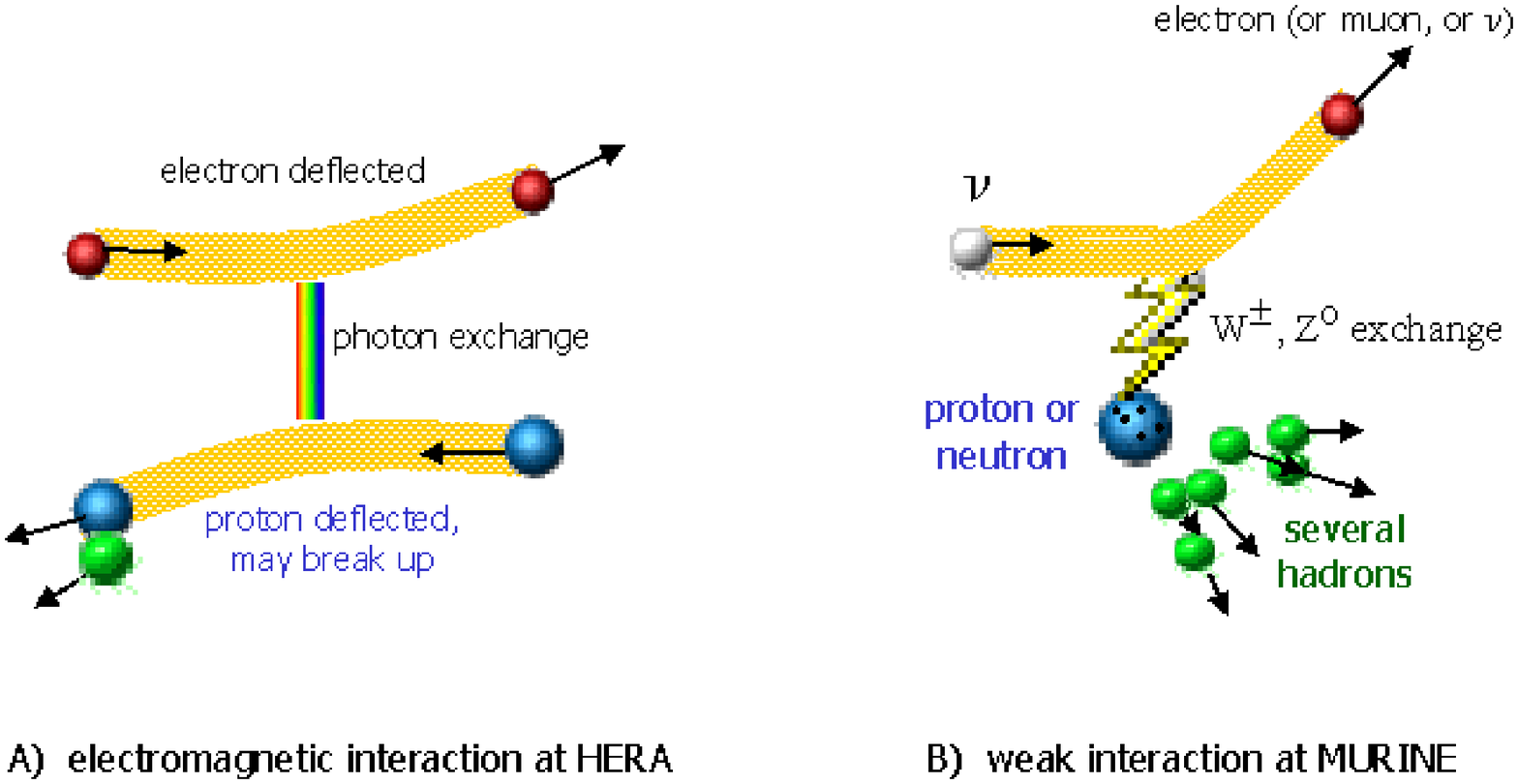}
\caption{
Conceptual illustration of A) the relatively soft electromagnetic
interactions, involving the exchange of a photon, that dominate
the event sample at HERA, and B) the much harder weak interactions
that will occur in Mighty MURINEs and that exist on the
``hard scattering tail'' at HERA.
}
\label{HERA_MURINE_interactions}
\end{figure}

  The HERA ep collider is a convenient reference point for
assessing the physics potential of Mighty MURINEs at the
highest energy scales. As indicated in figure~\ref{HERA_MURINE_interactions},
MURINEs have the same {\em weak} interactions as HERA while
avoiding the predominantly soft electromagnetic interactions that
dominate the HERA event trigger rates but are less interesting
because lower energy transfers probe physics at relatively
lower energy scales. The event samples at MURINEs will
correspond to the weak interactions in the
``hard scattering tail'' of the HERA event sample.

\begin{table}[ht!]
\caption{Comparison of a 50--100 TeV mighty MURINE with HERA for the
production of heavy particles. All of the numbers are order of magnitude
only, and are based on simple and approximate scalings from
reference~\cite{HERAref} rather than on detailed calculations.
}
\begin{tabular}{|ccc|}
\hline
Particle/Process & HERA & Mighty MURINE \\
        & (1 inverse femtobarn) & (1 inverse zeptobarn) \\
\hline
b quark: ${\rm c \rightarrow b}$
                            & O(10) events       & ${\rm O(10^6)}$ events \\
b quark: ${\rm u \rightarrow b}$
                            & O(1) event         & ${\rm O(10^5)}$ events \\
top quark                   & O(1) event         & ${\rm O(10^5)}$ events   \\
W, Z bosons                 & tens of events     & ${\rm O(10^5)}$ events ? \\
120 GeV SM Higgs            & O(1) event         & ${\rm O(10^5)}$ events   \\
exotica with small $\sigma$ & luminosity limited & luminosity OK!
\end{tabular}
\label{tab:physics}
\end{table}

  The HERA event samples involving weak interactions can be
compared with MURINEs at energies where the high energy tail of the
neutrino beam is comparable to the 314 GeV HERA center-of-mass
energy. The maximum neutrino energy is the muon beam energy,
which, according to equation~\ref{eq:MURINE_Ecom}, equals
the HERA CoM energy for $E_\mu = 53$ TeV. At this energy
or slightly above, very rough estimates of the event rates of
similar or identical processes can be simply transcribed
from HERA calculations after scaling by the $10^5$ luminosity
ratio shown in table~\ref{tab:EandL}.
Such a comparison is shown in table~\ref{tab:physics}.
A range of MURINE energies has been given, in deference to the
very approximate nature of the comparison. At the low end
($E_\mu=50$ TeV), the MURINE event rates will probably be lower than
the estimate, and the rates will normally be higher at the
high end ($E_\mu=100$ TeV).

 The standard model physics processes involving weak interactions
are the same in all cases except for the
production of W and Z bosons, where HERA has the advantage due to
processes involving photon exchange. The SM Higgs
has not been found at the time of writing, so the 120 GeV mass is
an example only.

 The first three processes in table~\ref{tab:physics}
have already been discussed in the preceding section.
To be realistic, at the event rates shown it is very doubtful that
W, Z and SM Higgs production could contribute anything useful beyond
collider studies, despite the the astounding neutrino beam parameters
and superior event reconstruction. Beyond this, possible exotic processes
at the 100 GeV scale or below provide the only substantial potential
for exciting discoveries. This motivation could become much
stronger in the near future if, for example, one of the current
leptoquark searches at HERA  returned a discovery. It is noted
that the leptoquarks produced at a Mighty MURINE might well be
different -- coupling to neutrinos rather than electrons -- and
so studies at MURINEs could potentially be complementary to
those at a future ep collider with a higher $\ECoM$ than HERA.

\section{Summary}
\label{sec:summary}

  The Mighty MURINE neutrino experiments that would come almost
for free at any future many-TeV muon collider could
improve on the pioneering advances from the previous MURINEs
that would have existed at lower energy muon colliders.
The most important improvements
might well be on the unique and important
measurements from previous lower energy MURINEs of
$\Vub$  and $\Vcb$, perhaps pushing the accuracy of both
measurements below 1\%. With total event statistics of a few times
$10^{11}$ events, the rare production of B's through flavor
changing neutral current interactions off valence d quarks
might provide one of the best indirect determinations of $\Vtd$.
More common channels for B production, particularly through
neutral current interactions, might also provide some capabilities
as a B factory with novel experimental strengths.

  Upon crossing the threshold for
top production, the even more interesting elements
$\Vtd$, $\Vts$ and $\Vtb$ could become successively
available to uniquely precise measurements at the highest energy
Mighty MURINEs.
The addition of any or all of these three precise measurements
would clearly
advance our knowledge of the CKM matrix to a level where small
perturbations from the Standard Model scenario could be searched for and,
if found, could be studied. MURINEs would then truly
play the central role in determining the CKM matrix parameters,
with the best measurements of the magnitudes of perhaps seven
of the nine elements (all but the two elements that are
currently best measured: $\Vud$ and $\Vus$) to add to the
phase information from various other experimental processes.

 If muon colliders ever reach the 100 TeV center of mass energy scale then
their neutrino experiments will attain a center of mass energy reach
comparable to the existing HERA ep collider, but at a luminosity
that might be perhaps 5 orders of magnitude higher. HERA then
becomes a convenient reference point for assessing their physics
capabilities. Despite the promise of impressive luminosities, none
of the standard model processes other than the CKM matrix appear to
offer the chance of competitive physics potential to studies of the
same processes at colliders.
Therefore, only i) an enlarged theoretical importance for
the CKM matrix or ii) the discovery, then or beforehand, of an exotic
process that is accessible to Mighty MURINEs, would give Mighty
MURINEs a chance for physics analyses of a comparable importance
to those at the colliders. Leptoquarks that couple to neutrinos
are the obvious candidate for such a new process.


\begin{references}
\bibitem{bjkthesispaper}  B.J. King,
    {\it Assessment of the Prospects for Muon Colliders},
    paper submitted in partial fulfillment of requirements
    for Ph.D., Columbia University, New York (1994),
    available from http://xxx.lanl.gov/ as {\bf physics/9907027}.
\bibitem{geer}  S. Geer, {\it Neutrino beams from Muon Storage Rings:
          Characteristics and Potential},  PRD 57, 6989 (1998).
\bibitem{hemc99nurad}  B.J. King,
   {\it Neutrino Radiation Challenges and Proposed Solutions for
   Many-TeV Muon Colliders}, these proceedings.
   Also available from~\hfill\break
   \verb|http://pubweb.bnl.gov/people/bking/heshop/hemc_papers.html|.
\bibitem{hemc99intro}
      B.J. King,
    {\it  Prospects for Colliders and Collider Physics to the
          1 PeV Energy Scale}, {\it ibid.}
\bibitem{numcbook}
   I.I. Bigi {\it et al.},
  ``The potential for High Rate Neutrino Physics at Muon
   Colliders and Other Muon Storage Rings'', in preparation
   for publication in Physics Reports.
\bibitem{quigg} See, for example, Chris Quigg,
    {\it Neutrino Interaction Cross Sections}, FERMILAB-Conf-97/158-T.
\bibitem{nufnal97} B.J. King,
    {\it Neutrino Physics at a Muon Collider},
    Proc. Workshop on Physics at the First Muon Collider
    and Front End of a Muon Collider, Fermilab, November 6-9, 1997,
    available from http://xxx.lanl.gov/ as {\bf hep-ex/9907033}.
\bibitem{SLD}  K. Abe {\it et al.},
    {Design and Performance of the SLD Vertex Detector: a 307 Mpixel
    Tracking System}, NIM A 400 (1997) 287-343.
\bibitem{Timcomm}  Private communication with Tim Bolton.
\bibitem{Bigi}  Ikaros Bigi pointed out the theoretical importance of
    this B decay process to the author and suggested considering the
    experimental possibilities to study it at MURINEs.
\bibitem{HERA_CKM}  R.M. Godbole {\it et al.},
    {\it The Kobayashi-Maskawa Matrix at HERA}, Proceedings of the HERA
    Workshop, Hamburg, October 12-14, 1987, ed. R.D. Peccei, publ. DESY;
    A. Ali {\it et al.},
    {\it Heavy Quark Physics at HERA}, {\it ibid.}
\bibitem{Vcdmeas}  H. Abramowicz {\it et al.}, Z. Phys. {\bf C21}, 27 (1982);
       S.A. Rabinowicz {\it et al.}, Phys. Rev. Lett. {\bf 70}, 134 (1993);
       A.O. Bazarko {\it et al.}, Z. Phys. {\bf C65}, 189 (1995).
\bibitem{PDG_CKM}  C. Caso {\it et al.}, {\it The Review of Particle
Physics}, The European Physical Journal C3 (1998) 1 
    and 1999 off-year partial update for the 2000 edition available on 
    the PDG WWW pages (URL: $http://pdg.lbl.gov/$).
\bibitem{jh99}  B.J. King,
    {\it High Rate Physics at Neutrino Factories},
    Submitted to Proc. 23rd Johns Hopkins Workshop on Current Problems
    in Particle Theory, "Neutrinos in the Next Millenium",
    Johns Hopkins University, Baltimore MD, June 10-12, 1999,
    available from http://xxx.lanl.gov/ as {\bf hep-ex/9911008}.
\bibitem{E787}
    S. Adler {\it et al.},
    {\it Evidence for the Decay
     $K^- \rightarrow \pi^- \nu \overline{\nu}$  }
     Phys. Rev. Lett. 79 (1997) 2204-2207, 
     hep-ex/9708031;
    S. Adler {\it et al.},
    {\it Further Search for the Decay
    $K^- \rightarrow \pi^- \nu \overline{\nu}$  }
    hep-ex/0002015.
\bibitem{Littenberg}  Private communication with L.S. Littenberg.
\bibitem{Zimmermann}  F. Zimmermann,
    {\it Final Focus Challenges for Muon Colliders at Highest Energies},
          these proceedings. Also available from~\hfill\break 
          \verb|http://pubweb.bnl.gov/people/bking/heshop/hemc_papers.html|.
\bibitem{HERAref}  Proceedings of the HERA
    Workshop, Hamburg, October 12-14, 1987, ed. R.D. Peccei, publ. DESY.
\end{references}
\end{document}